\documentclass[preprint,12pt]{elsarticle}

\usepackage{amssymb}

\journal{Scientific Reports}

\begin{document}

\begin{frontmatter}



\title{A universal constant for dark matter-baryon interplay}


\author{Man Ho Chan}

\address{Department of Science and Environmental Studies, The Education University of Hong Kong \\ 
Tai Po, New Territories, Hong Kong, China}

\ead{chanmh@eduhk.hk}

\begin{abstract}
Recent studies point out that there exists some rough scaling relations for dark matter and some tight connections between dark matter and baryons. However, most of the relations and tight connections can only be found in galaxies, but not in galaxy clusters. In this article, we consider a new expression that can characterize the properties of dark matter-baryon interplay for both galactic and galaxy cluster scales. By using the archival observational data of galaxies and galaxy clusters, we show that the value $K=\bar{n}_D \bar{n}_Br_oV/v^4$ is almost a constant and scale independent within the optical radius $r_o$, where $\bar{n}_D$ is the average dark matter number density, $\bar{n}_B$ is the average baryon number density, $v$ is the characteristic velocity and $V$ is the interacting volume. This would be the first universal relation between dark matter and baryons on both galactic and galaxy cluster scales. We anticipate this result to be a starting point to explain the small-scale problem and the scaling relations for dark matter in galaxies. The constant $K$ discovered may reveal some underlying global interaction between dark matter and baryons.
\end{abstract}

\begin{keyword}
Dark Matter
\end{keyword}

\end{frontmatter}


\section{Introduction}
It is commonly believed that dark matter particles exist in galaxies and galaxy clusters. However, we know very little about the properties of dark matter. Some studies point out that there exists some rough scaling relations for dark matter. For example, the product of the central density and the characteristic scale radius of dark matter is almost a constant \cite{Donato,Burkert,Karukes}. However, these relations can only apply for galaxies, but not for galaxy clusters \cite{Chan}. Besides, based on the assumption of collisionless behavior of dark matter particles, computer simulations show that dark matter density would follow the Navarro-Frenk-White (NFW) profile \cite{Navarro}, which is usually regarded as a universal density profile for dark matter. However, this profile only gives good agreement with the data of galaxy clusters \cite{Viola} and some galaxies \cite{Iocco,Sofue}, but not for dwarf galaxies \cite{Zackrisson,deBlok}. Therefore, it does not have any universal relation for dark matter which is true for both galactic and galaxy cluster scales. Some suggestions have been proposed to explain the discrepancies between the results of observations and computer simulations, such as baryonic feedback from supernovae and star formation \cite{Governato,Pontzen}, and self-interacting dark matter model \cite{Spergel,Burkert2,Koda}. However, it is doubtful that the baryonic matter can rigorously control the dark matter density profile in many dwarf galaxies \cite{Merritt}. Also, recent observations give stringent constraints on self-interacting cross section of dark matter such that some tension still exists between predictions and observations (for a review of this problem, see \cite{Tulin}).

On the other hand, there are some interesting relations which relate dynamical mass (or dark matter halo mass) with baryonic mass, such as the Tully-Fisher relation \cite{Tully,McGaugh2,Lelli2}, the mass-discrepancy-acceleration relation \cite{McGaugh3}, the central-surface-densities relation \cite{Lelli} and the radial acceleration relation \cite{McGaugh}. These relations indicate some specific connection between dark matter and baryonic matter, which reveal some forms of interplay between dark matter and baryons. Recent 21-cm radio observations indicate a stronger absorption signal than expected at cosmic dawn (the EDGES result) \cite{Bowman}, which may reveal a non-zero interaction between dark matter and baryons \cite{Barkana}. However, further studies show that the EDGES result is unlikely to be solved by a dominant component of millicharged dark matter, allowing only to consist of 1\% of the total dark matter \cite{Barkana2}. Some studies even show that only less than 0.4\% fraction is allowed for millicharged dark matter \cite{Kovetz,Boddy}. Therefore, the millicharged dark matter model cannot play a significant role on galactic and cluster scales. 

In this article, we show that there exists a universal relation between dark matter and baryons on both galactic and galaxy cluster scales. The universal relation can be described by a constant which characterizes the properties of dark matter-baryon interplay. The value of the constant is almost universal for both galactic and galaxy cluster scales. This constant may reveal some underlying global properties of dark matter-baryon interaction or self-organization, which may also be used to explain the `core-cusp problem' \cite{deBlok}, `the mass-discrepancy-acceleration relation' \cite{McGaugh} and the `halo-disk conspiracy problem' \cite{Battaner}.

\section{Method}
The effect of dark matter-baryon interplay inside an `interacting region' would depend on the average dark matter number density $\bar{n}_D$, average baryon number density $\bar{n}_B$ and the volume of the interacting region $V$ which depends on the optical radius $r_o$. Since most of the galaxies are disk-dominated at $r_o$, the enclosed interacting volume would be a disk-like thin cylinder with radius $r_o$ (see Fig.~1). Consider the following product (unit: cm$^{-2}$)
\begin{equation}
K= \frac{\bar{\rho}_D \bar{\rho}_B}{m_Dm_B} \left( \frac{v}{c} \right)^{-4}r_oV,
\end{equation} 
where $\bar{\rho}_D=m_D \bar{n}_D$, $\bar{\rho}_B=m_B \bar{n}_B$, $v$ is the characteristic velocity, $m_D$ is the dark matter mass and $m_B$ is the average mass of a baryonic particle. Here, we have assumed that all dark matter can interact with baryons. The potential interaction can be gravitational only, or including some other forms of interactions such as scattering.  

\subsection{The constant for galaxies}
For spiral galaxies, the Tully-Fisher relation relates the total baryonic mass with velocity as $M_B=(47 \pm 6)M_{\odot}$km$^{-4}$ s$^{-4}$ $v'^4$ \cite{McGaugh2}, where $v'$ is the asymptotic circular velocity. Since most of the galactic rotation curves rise to a constant $v'$ quickly within $r_o$, we assume that the value of $v'$ is equal to the characteristic velocity $v$ of a galaxy. 

Besides, the optical radius $r_o$ is closely related to the scale radius of dark matter $r_s$. By using the data of the local volume catalog (LVC) dwarf galaxies \cite{Karukes,Salucci}, a very strong correlation between $r_s$ and $r_o$ is found (see Fig.~2). Although the relation is not a perfect linear relation, their ratio is close to a constant $r_s=(2.59 \pm 0.55)r_o$. Since $M_B=\bar{\rho}_BV=(47 \pm 6)M_{\odot}$km$^{-4}$ s$^{-4}$ $v^4$, the term $v$ in Eq.~(1) is canceled. Furthermore, putting $r_s=(2.59 \pm 0.55)r_o$ into Eq.~(1) and taking $m_B=1.2m_p$, where $m_p$ is the proton mass, we can calculate the value and the uncertainty of $K$:
\begin{equation}
K=1.7^{+0.8}_{-0.4} \times 10^{102} \left( \frac{\bar{\rho}_Dr_s}{100M_{\odot}\rm pc^{-2}} \right) \left( \frac{m_D}{\rm GeV} \right)^{-1}~{\rm cm^{-2}}.
\end{equation}

On the other hand, the product $K$ can also be understood as follow. Since $v=\sqrt{GM_D(r)/r}$, where $M_D(r)$ is the enclosed dynamical mass of a galaxy, we have $v^4 \propto M_D^2r_o^{-2}$. Assuming $M_D(r_o) \propto \bar{\rho}_Dr_o^3$, we get
\begin{equation}
K \propto \frac{M_B}{M_D(r_o)}.
\end{equation}
Therefore, $K$ is directly proportional to the ratio of the baryonic mass and the enclosed dynamical mass at radius $r_o$. Note that this ratio is not necessarily equal to the cosmic baryon fraction $f=0.156 \pm 0.003$ \cite{Planck}. For different sizes and types of galaxies, it is possible to have entirely different ratios at radii $r_o$ (see the discussion below).

Furthermore, the above relation is obtained by using the Tully-Fisher relation, which is formulated based on the data of spiral galaxies. For elliptical galaxies, many of them satisfy another relation called the Faber-Jackson relation \cite{Faber}. This relation indicates a close relationship between the luminosity $L$ and the stellar velocity dispersion $v_b$ of elliptical galaxies: $L \propto v_b^4$, which is similar to the form of the Tully-Fisher relation. Recent studies show that the data of some elliptical galaxies also fall on the spiral baryonic Tully-Fisher relation if one assumes a certain value of mass-to-luminosity ratio \cite{Heijer}. Therefore, the above result may also be applicable for elliptical galaxies.

\subsection{The constant for galaxy clusters}
The above deduction method can only be applied for galaxies. It is because there is no Tully-Fisher relation for galaxy clusters. Also, the product $\bar{\rho}_Dr_s$ is not a constant for galaxy clusters \cite{Chan}. Fortunately, we can relate the velocity of baryons $v'$ by the temperature $T$ of hot gas in galaxy clusters, which can be given by Virial relation $v \approx \sqrt{3kT/m_p}$. The baryonic component in a large galaxy cluster is dominated by the spherical hot gas halo within the core radius of the hot gas $r_c$. Therefore, we consider the spherical interacting region inside $r=r_o=r_c$ so that $V=4\pi r_c^3/3$. Besides, the hot gas is in hydrostatic equilibrium so that the total mass can be determined \cite{Chen}:
\begin{equation}
\frac{dP}{dr}=-\frac{GM(r)\rho_B}{r^2},
\end{equation}
where $P=\rho_BkT/(\mu m_p)$ is the pressure of hot gas, $\mu=0.61$ is the molecular weight and $M(r)$ is the total enclosed mass. Recent studies show that the x-ray hydrostatic mass measurements are remarkably robust and method-independent \cite{Bartalucci}. The overall percentage error of mass estimation is about 10-30\% if we assume that $T$ is a constant \cite{Reiprich,Chen}. The hot gas surface brightness profile can be determined by x-ray observations. It is usually described by a $\beta$-model \cite{Chen}: 
\begin{equation}
S(r)=S_0 \left[1+ \left(\frac{r}{r_c} \right)^2 \right]^{-3\beta+1/2},
\end{equation}
where $S_0$ is the central surface brightness, $r_c$ is the core radius and $\beta$ is a fitted parameter. These parameters can be used to construct the density profile of hot gas \cite{Reiprich}:
\begin{equation}
\rho_B=\rho_{B0} \left(1+ \frac{r^2}{r_c^2} \right)^{-3 \beta/2},
\end{equation}
where $\rho_{B0}$ is the central baryon density. Combining Eq.~(4) and Eq.~(6), we get
\begin{equation}
M(r)=\frac{3kT\beta r^3}{\mu m_pG(r_c^2+r^2)}.
\end{equation}
The total mass density can be obtained by 
\begin{equation}
\rho_t=\frac{1}{4\pi r^2} \frac{dM(r)}{dr}=\frac{3kT\beta}{4\pi G \mu m_p} \left[ \frac{3r_c^2+r^2}{(r_c^2+r^2)^2} \right].
\end{equation}
Taking $r\rightarrow 0$, we can get the central total mass density $\rho_{t0}$. Since dark matter dominates the mass in a galaxy cluster, the central dark matter density is close to the central mass density:
\begin{equation}
\rho_{D0} \approx \rho_{t0}=\frac{9\beta kT}{4\pi \mu m_pGr_c^2}.
\end{equation}
Note that for most of the galaxy clusters, the density profiles are close to the NFW profile (a cuspy profile). Here, the density $\rho_{D0}$ represents the average dark matter density for $r \le r_c$ (i.e. $\bar{\rho}_D \approx \rho_{D0}$). Using Eq.~(9), the value of $\rho_{D0}$ for each galaxy cluster can be calculated by the parameters $T$, $\beta$ and $r_c$, including their uncertainties.

Putting all the above relations to Eq.~(1), the term $v$ and $r_c$ would be cancelled naturally. Therefore, we get
\begin{equation}
K= 1.4 \times 10^{103} \left( \frac{\bar{\rho}_B}{\bar{\rho}_D} \right) \left(\frac{m_D}{\rm GeV} \right)^{-1} \left( \frac{\beta}{0.65} \right)^2~{\rm cm^{-2}}.
\end{equation}

\section{Result}
\subsection{For galaxies}
In many dark matter-dominated galaxies, the mass density of dark matter for $r \le r_s$ is close to constant (cored structure: $\bar{\rho}_D \approx \rho_{D0}$) \cite{deBlok,Contenta}. Also, recent analyses suggest that the central dark matter column density (the product of the central dark matter mass density and the scale radius) is almost a constant. Early analysis using $\sim 1000$ galaxies gives $\rho_{D0}r_s=141^{+82}_{-52}M_{\odot}$ pc$^{-2}$ \cite{Donato} while later analysis using Milky Way spheroidal dwarf galaxies gives $\rho_{D0}r_s=75^{+85}_{-45}M_{\odot}$ pc$^{-2}$ \cite{Burkert}. After that, using a sample of the LVC dwarf galaxies, the resultant central surface mass density gives $\rho_{D0}r_s \sim 100M_{\odot}$ pc$^{-2}$ \cite{Karukes}, which is consistent with the previous studies. By combining the data of the Milky Way spheroidal dwarf galaxies \cite{Burkert} and the sample of disc galaxies in \cite{Karukes}, we get $\bar{\rho}_D \propto r_s^{-0.97 \pm 0.14}$ and $\bar{\rho}_Dr_s=95^{+70}_{-40}M_{\odot}$ pc$^{-2}$ (see Fig.~3). Therefore, using this result, we get 
\begin{equation}
K=1.7^{+2.4}_{-1.1} \times 10^{102} \left( \frac{m_D}{\rm GeV} \right)^{-1}~{\rm cm^{-2}}.
\end{equation}

The narrow range of $K$ (or the ratio of baryonic mass to the enclosed dynamical mass) for galaxies has been known for a decade. The almost constant central dark matter column density for a wide range of galaxies \cite{Donato,Burkert,Karukes} and the constant luminous-to-dark matter ratio within one halo scale-length \cite{Gentile} suggest the narrow range of $K$. Therefore, the almost constant value of $K$ for galaxies is expected. Note that some studies have challenged the non-universality of the central dark matter column density in galaxies \cite{Saburova,Rodrigues}. Nevertheless, this is a controversial issue and we may assume that this is approximately true in galaxies to deduce the implications based on this relation.
 
\subsection{For galaxy clusters}
By using the x-ray data of 64 large galaxy clusters (with $r_c \ge 100$ kpc) \cite{Chen}, we get $\rho_{D0} \propto \rho_{B0}^{1.00 \pm 0.06}$ (see Fig.~4), where $\rho_{B0} \approx \bar{\rho}_B$ is the central mass density of hot gas. It means that the ratio $\rho_{D0}/\rho_{B0} \approx 8.70 \pm 3.42$ is roughly a constant. This correlation has not been discovered and discussed in previous studies. Although the average dark matter density to average baryonic density ratio for a galaxy cluster should be close to $f^{-1} \approx 6.41$, the ratio at the central region for all galaxy clusters may not be a constant and equal to $f^{-1}$ because the density profiles for dark matter and hot gas are entirely different (see the discussion below). This approximately constant ratio of $\rho_{D0}$ to $\rho_{B0}$ at $r_c$ may reveal some global interplay between dark matter and baryons.

Although different galaxy clusters have different values of $\beta$ \cite{Chen}, we approximate the distribution of $\beta$ by a Gaussian function with an average value $\beta=0.65$ and a dispersion of 0.13 (see Fig.~5). Putting the relation of $\rho_{D0}$ and $\rho_{B0}$ to Eq.~(10) and using $\beta=0.65 \pm 0.13$, we get a scale invariant constant for galaxy clusters:
\begin{equation}
K \approx 1.6^{+2.3}_{-0.9} \times 10^{102} \left( \frac{m_D}{\rm GeV} \right)^{-1}~{\rm cm^{-2}}, 
\end{equation}
where $\bar{\rho}_D/\bar{\rho}_B \approx \rho_{D0}/\rho_{B0}$. We can see that the values of $K$ for galaxies and galaxy clusters give excellent agreement with each other, within a factor of 2-3. This constant can be regarded as a universal constant for dark matter-baryon interplay because it is independent of scale. 

Here, note that the x-ray data used have assumed the hubble parameter $h=0.5$ \cite{Chen}. To match the current observed hubble parameter $h=0.68$ \cite{Planck}, we have re-scaled the parameters $r_c$ and $\rho_{B0}$. Besides, using the data of large galaxy clusters can minimize the errors because the uncertainties of the central hot gas temperature significantly affect the mass determination of small galaxy clusters. Therefore, we only choose the large galaxy clusters (with $r_c \ge 100$ kpc) to perform the analysis. 

\subsection{The relation between the constant and the cosmic baryon fraction}
As mentioned above, the value of $K$ is directly proportional to the ratio $M_B/M_D(r_o)$ (for galaxies) or $\rho_{B0}/\rho_{D0}$ (for galaxy clusters). One may claim that it is not surprising to have a similar constant value of $K$ for galaxies and galaxy clusters because $M_B/M_D(r_o)$ or $\rho_{B0}/\rho_{D0}$ should be close to the cosmic baryon fraction $f=0.156 \pm 0.003$, which should be almost a constant for various structures. However, previous studies show that the baryon fractions in various structures are entirely different and the ratios quite depend on the positions of the structures \cite{Gonzalez,McGaugh4}. In Fig.~6, we follow the study in \cite{McGaugh4} and show the corresponding baryon fractions for different structures (from dwarf galaxies to galaxy clusters). We can see that the ratio can be ranging from $10^{-3}$ to $10^{-1}$, which spans 3 orders of magnitude.

If we assume the NFW profile for dark matter distribution in galaxy clusters and fix the baryon fraction $f=M_B/(M_B+M_D)=0.156$ at a large radius $r=1$ Mpc, we can predict the baryon fraction at the characteristic radius $r_s$ for various total dynamical mass using hydrostatic equilibrium. In Fig.~7, we can see that the baryon fraction at $r_s=100$ kpc ranges from $10^{-3}$ to $10^{-1}$ (not close to $f=0.156 \pm 0.003$) for just a short range of dynamical mass if we consider a fix hot gas temperature ($T=5$ keV) and scale radius of dark matter $r_s=100$ kpc. For various temperature of hot gas and scale radii of different galaxy clusters, the variation of the baryon fraction can be much larger. This large variation can also be seen in galaxies by examining the radial acceleration relation \cite{McGaugh}. Therefore, theoretically, the baryon-dark matter density ratio is not necessarily a constant for various structures, but possibly spans at least 2 orders of magnitude. Furthermore, the characteristic radii of the galaxies ($\sim 1-10$ kpc) and galaxy clusters considered ($\sim 100-500$) kpc are different. There is no reason why the baryon fractions for galaxies and galaxy clusters are close to each other at different characteristic radii. Nevertheless, our results show that this ratio is almost constant at the characteristic radii $r_o$ or $r_c$. Based on the result of Fig.~7, it seems that gravitational interaction between dark matter and baryonic matter alone is not enough to provide a satisfactory explanation for the constant value of $K$. This may reveal the existence of some interplay or self-organizing processes between baryons and dark matter particles within the `interacting region'.

\section{Discussion}
Previous studies have claimed some relations for dark matter. For example, as mentioned above, the central dark matter column density $\rho_{D0}r_s$ is almost a constant for a wide range of galaxies \cite{Karukes}. Some explanations have been given to account for this relation, such as the self-interacting dark matter model \cite{Chan2,Tulin}. However, this relation is not correct for galaxy clusters \cite{Chan}. Therefore, such scaling relation is not universal at all. Besides, the alleged universal NFW density profile for dark matter only gives good agreement with the data of galaxy clusters \cite{Viola} and some galaxies \cite{Iocco,Sofue}, but not for dwarf galaxies \cite{Zackrisson,deBlok}. Therefore, it does not have any universal relation for dark matter which is true for both galactic and galaxy cluster scales. Our result, $K$ is a constant, would be the first universal relation of dark matter for both galactic and galaxy cluster scales. Although this constant is proportional to the ratio of baryon fraction in galaxies and galaxy clusters at the optical radii or core radii, it is not necessarily a constant. As shown in Fig.~6, the baryon fractions can span 3 orders of magnitude variation for different structures. Therefore, our results suggest that some interplay between baryons and dark matter particles may exist so that the constant $K$ or the baryon fraction is almost a constant for galaxies and galaxy clusters within the `interacting region'.

Note that the choice of the characteristic radius ($r_o$ or $r_c$) in this analysis is not arbitrary. The overlapping regions of dark matter and baryons for galaxies and galaxy clusters are characterized by the optical radii $r_o$ and the core radii $r_c$ respectively. The size of the interacting volume $V$ also depends on $r_o$ or $r_c$. If we consider the scale radius of dark matter $r_s$ as the characteristic radius, the volume considered would be much larger, which includes some non-interacting regions. Therefore, it is justified to use $r_o$ or $r_c$ as the characteristic radius for this analysis.

Although the range of baryon fraction can be very large, in this analysis, we mainly examine the value of $K$ (proportional to the central baryon fraction) within the `interacting region' (i.e. within $r_o$ or $r_c$). As mentioned above, the narrow range of $K$ in the central regions of galaxies has been known for a decade \cite{Gentile}. However, the narrow range of $K$ in the core regions of galaxy clusters has not been discussed extensively. Many studies only focus on the total baryon fraction for the whole galaxy clusters \cite{Gonzalez,Lagana}, not for the central regions. Some studies have examined the central baryon fraction of some galaxy clusters \cite{Allen,Pratt}. However, no extensive discussion about the range of the central baryon fraction has been made. Besides, some discrepancies exist between numerical simulations and observational data for galaxy clusters near the central regions. Previous studies using computer simulations suggest a very wide range of baryon fraction for different radii \cite{Young}. However, the simulated baryon fraction is somewhat smaller than that of the observed data \cite{Allen,Pratt,Young}. For $r_c=100-500$ kpc, the corresponding baryon fraction should be $\sim 0.02-0.1$ based on the simulated results \cite{Young}. This gives $\rho_{D0}/\rho_{B0} \sim 10-50$. Nevertheless, our results suggest a narrower and smaller ratio $\rho_{D0}/\rho_{B0}=8.70 \pm 3.42$ (or a larger value of $K$) compared with the simulated results, which generally agree with the observed ranges of the central baryon fraction in \cite{Allen,Pratt}. We explicitly analyze the baryon fraction distribution at core radii of galaxy clusters and quantify the results to get the value of $K$. It is surprising to see the value of $K$ for galaxy clusters being close to that for galaxies within the `interacting region', which has not been discussed and discovered in the early works. Therefore, our analysis gives some new hints to investigate the alleged interplay between dark matter and baryons.

It has been suggested that dark matter-baryon interaction can explain the observed `mass-discrepancy-acceleration relation' \cite{Famaey}. This relation is closely related to the `Tully-Fisher relation' \cite{Wheeler}. Some recent studies also discover some correlations between dark matter and baryonic component in galaxy clusters \cite{DelPopolo}. We anticipate that the global properties of dark matter-baryon interaction can give a holistic picture to account for these relations, as well as the `halo-disk conspiracy problem' \cite{Battaner}. Since the constant $K$ links up the baryon fractions with some of the scaling relations in galaxies, it may provide a more fundamental ground to account for the constant dark matter column density (or $\rho_{D0} \propto r_s^{-1}$) and the Tully-Fisher relation. Note that the dimension of the constant $K$ is cm$^{-2}$, which is the dimension of 1/cross section. This constant may be closely related to the cross section of the interplay between dark matter and baryons.

It seems that gravitational force only cannot provide a satisfactory account for this interplay. Since baryonic matter is self-interacting while dark matter particles are believed to be collisionless, the density profiles for baryonic matter and dark matter would be different. There is no reason why the values of $K$ (or the baryon fraction) within different characteristic radii are close to each other. Some suggest that baryonic feedback mechanisms can account for the core formation in galaxies \cite{Pontzen}. If baryonic feedback can have some self-organizing properties, it may be able to explain the universal nature of $K$. 

\begin{figure}
\vskip 10mm
 \includegraphics[width=120mm]{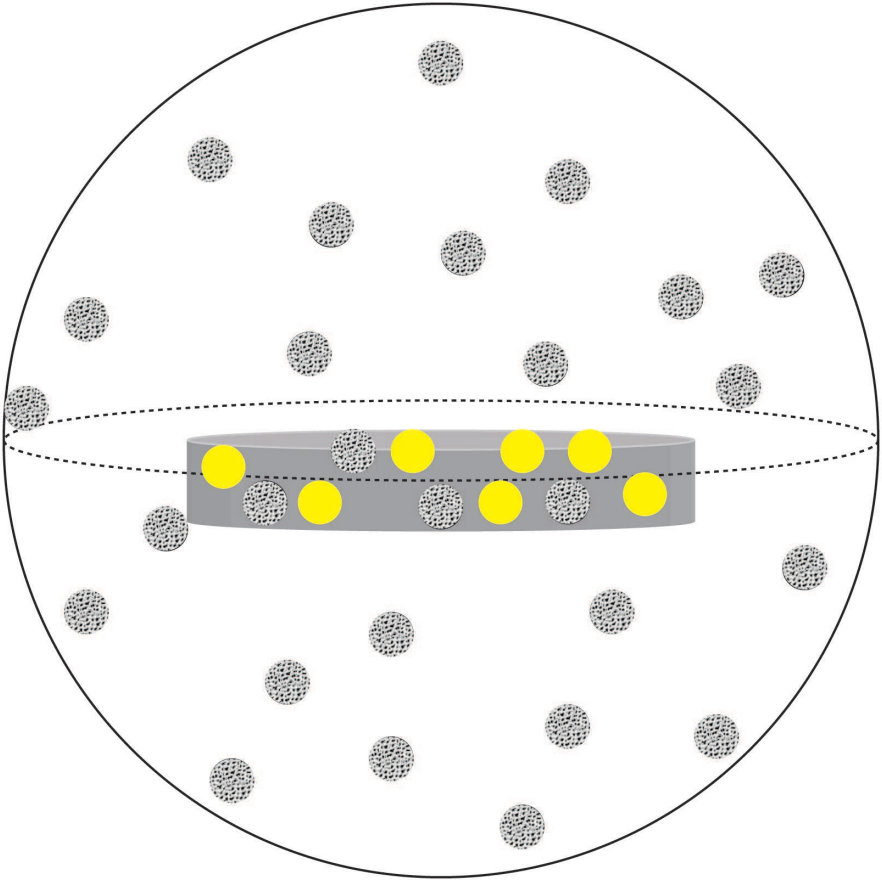}
 \caption{The schematic diagram of the interacting region in a dwarf galaxy. The large sphere is the dark matter core halo with radius $r=r_s$. The grey thin cylinder with base radius $r_o$ indicates the interacting volume $V$ (the effective volume of baryonic matter in a galaxy). The small yellow and black shaded circles represent the baryonic particles and dark matter particles respectively.}
\vskip 10mm
\end{figure}

\begin{figure}
\vskip 10mm
 \includegraphics[width=140mm]{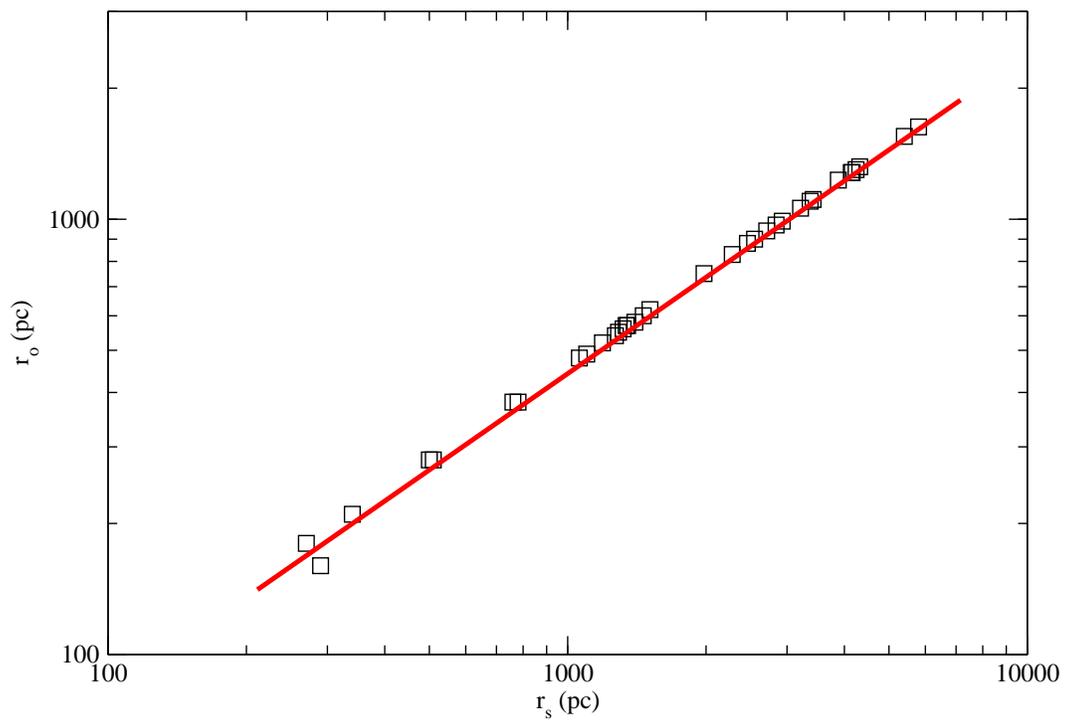}
 \caption{The graph of $r_o$ versus $r_s$. The data (indicated by squares) are taken from \cite{Karukes}. The red line indicates the tight correlation between the two quantities.}
\vskip 10mm
\end{figure}

\begin{figure}
\vskip 10mm
 \includegraphics[width=140mm]{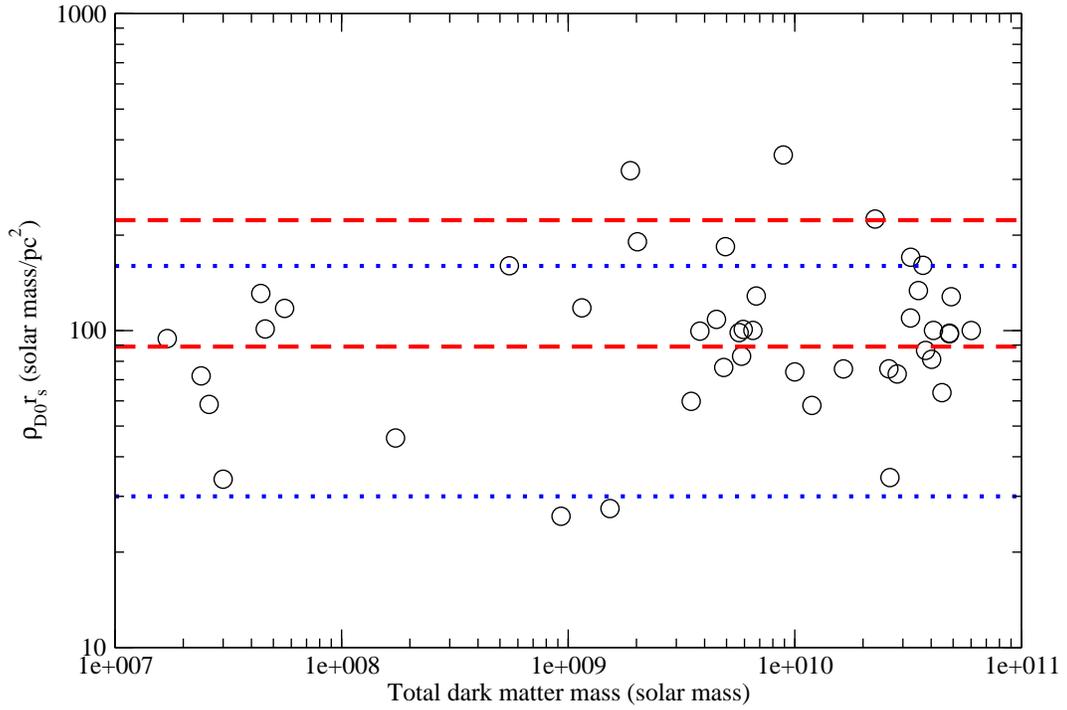}
 \caption{The central dark matter column density (in $M_{\odot}$ pc$^{-2}$) as a function of total galactic mass of dark matter. The data (indicated by circles) are taken from \cite{Burkert} and \cite{Karukes}. The regions bounded by the red dashed lines and bounded by the blue dotted lines are the ranges of the central dark matter column density obtained in \cite{Donato} and \cite{Burkert} respectively.}
\vskip 10mm
\end{figure}

\begin{figure}
\vskip 10mm
 \includegraphics[width=140mm]{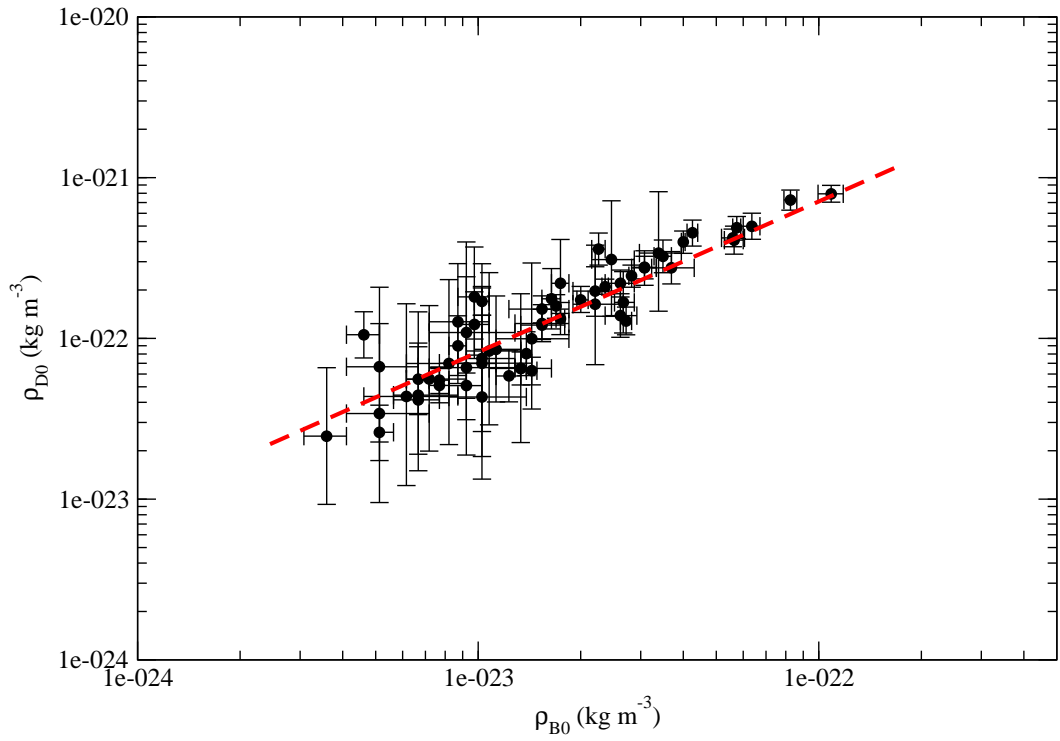}
 \caption{The graph of $\rho_{D0}$ versus $\rho_{B0}$ for 64 galaxy clusters (data are taken from \cite{Chen}). The red dashed line indicates the slope of the fit: $1.00\pm 0.06$.}
\vskip 10mm
\end{figure}

\begin{figure}
\vskip 10mm
 \includegraphics[width=140mm]{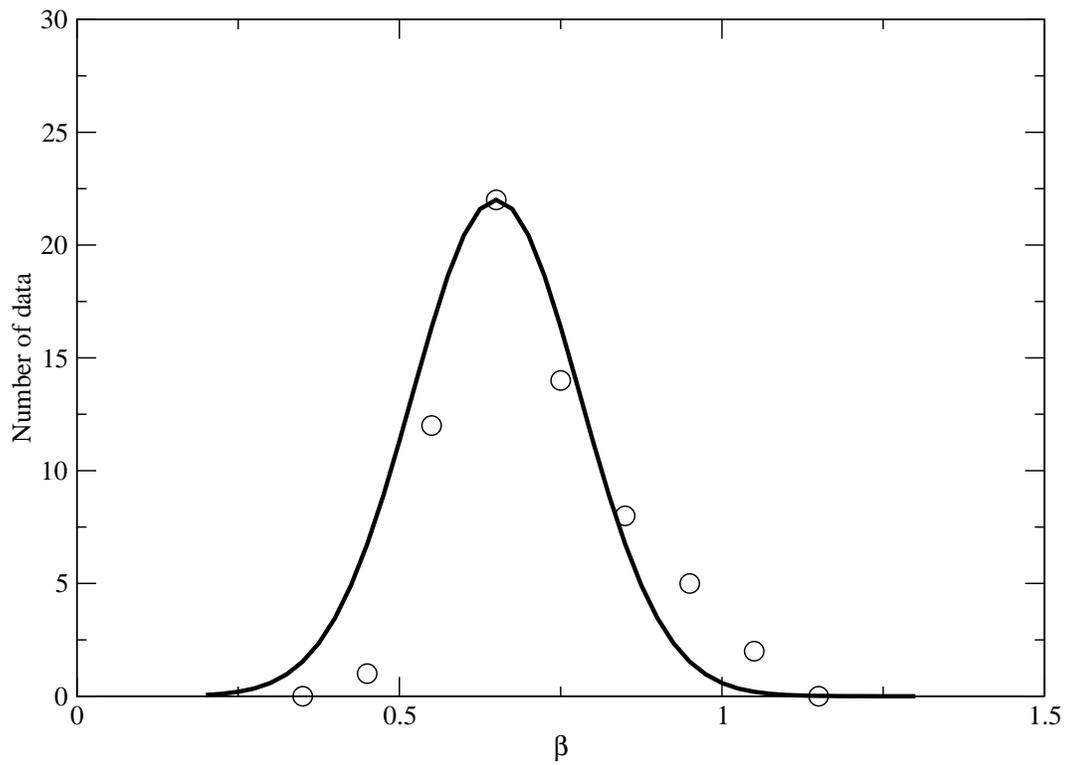}
 \caption{The distribution of $\beta$ for the 64 galaxy clusters (data are taken from \cite{Chen}). The black line indicates the Gaussian fit with $\beta=0.65\pm 0.13$.}
\vskip 10mm
\end{figure}

\begin{figure}
\vskip 10mm
 \includegraphics[width=140mm]{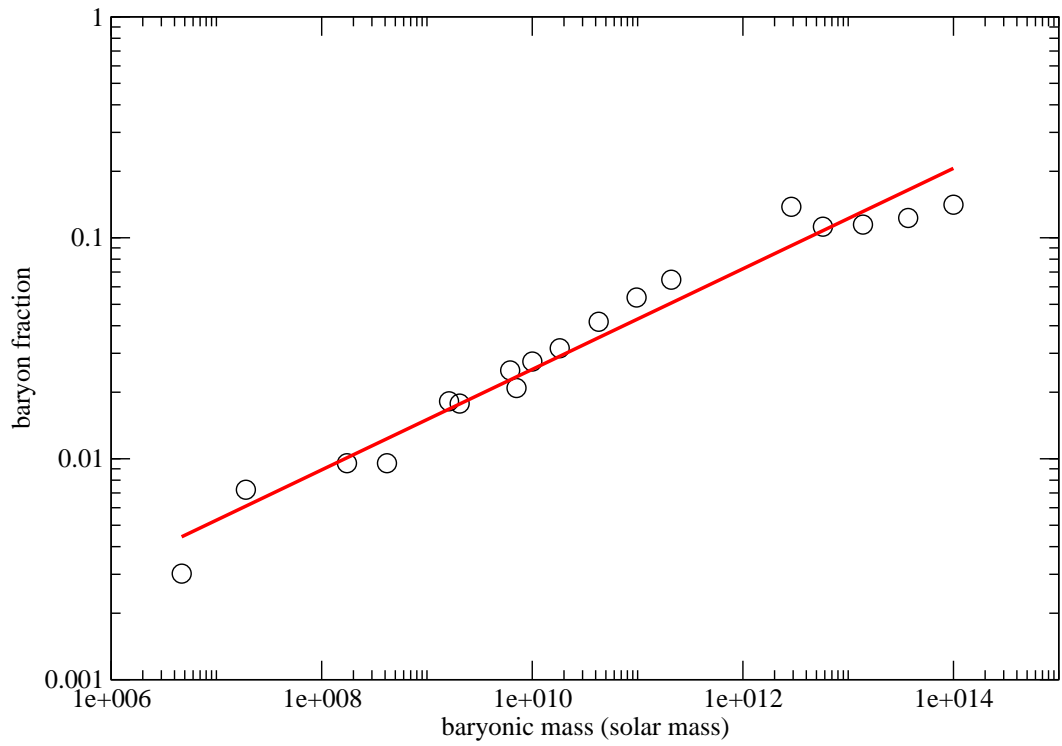}
 \caption{The graph of baryon fraction versus total baryonic mass for different structures (from dwarf galaxies to galaxy clusters) (data are taken from \cite{McGaugh4}). The red line indicates the correlation $f \propto M_B^{0.23}$.}
\vskip 10mm
\end{figure}

\begin{figure}
\vskip 10mm
 \includegraphics[width=140mm]{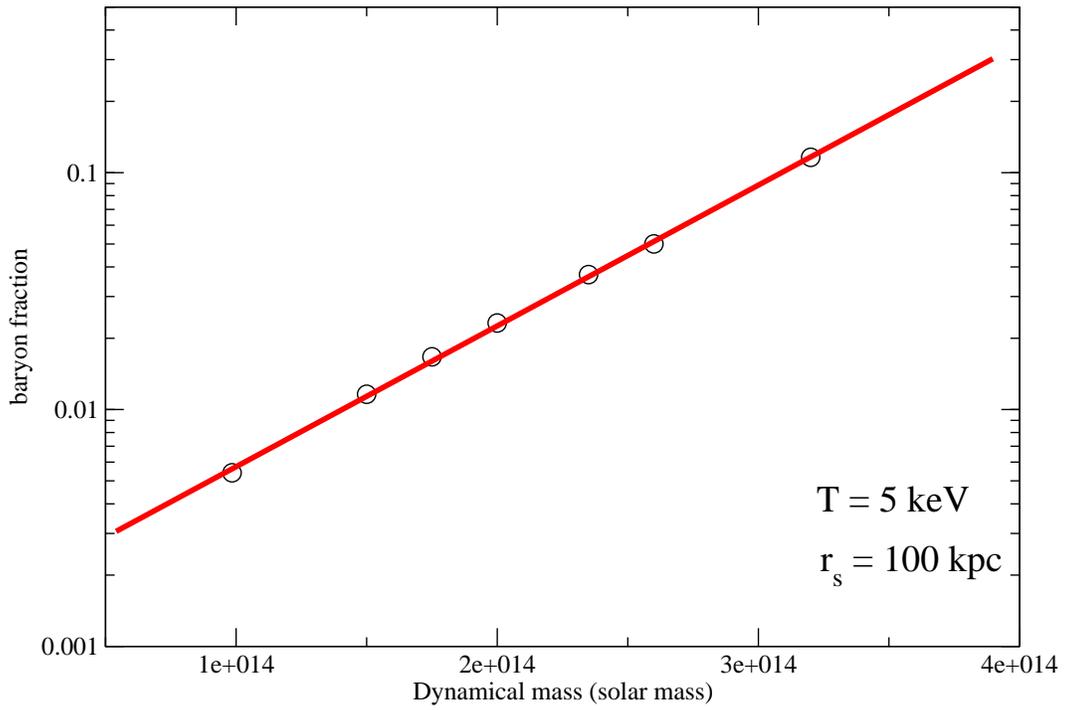}
 \caption{The graph of baryon fraction within $r_s$ versus total dynamical mass for galaxy clusters, assuming NFW density profile for dark matter and in hydrostatic equilibrium with hot gas. Here, we fix the hot gas temperature and scale radius of dark matter as 5 keV and $r_s=100$ kpc respectively.}
\vskip 10mm
\end{figure}

\section{acknowledgements}
The work described in this paper was partially supported by a grant from the Research Grants Council of the Hong Kong Special Administrative Region, China (Project No. EdUHK 28300518).

\section{Author contributions}
Design of the study, analysis of the results and writing of the manuscript were performed by Man Ho Chan.

\section{Data availability}
The datasets generated during and/or analysed during current study are available from the corresponding author on reasonable request.

\section{Competing interests}
The author declares no competing interests.





\end{document}